\documentclass[12pt]{iopart}
\usepackage{iopams}
\newcommand{\be} {\begin {equation}}
\newcommand {\ee} {\end {equation}}
\newcommand{\bea} {\begin {eqnarray}}
\newcommand {\eea} {\end {eqnarray}}
\newcommand{\ba}{\begin{array} } 
\newcommand{\ea}{\end{array}}
\newcommand {\ch} {\mathcal{H}}

\newcommand {\cl} {\mathcal{L}}
\newcommand{\nn}{\nonumber\\ }
\newtheorem{theorem}{Theorem}

\begin{document}

\title{A nonseparable quantum superintegrable system in 2D real {E}uclidean space. }

\author{Sarah Post$^1$, Pavel Winternitz$^{1,2}$}
\address{$^1$Centre de recherches math\'ematiques,
Universit\'e de Montr\'eal, \\C.P. 6128 succ. Centre-Ville, Montr\'eal (QC) H3C 3J7, Canada }
\address{$^2$ D\'epartement de math\'ematiques et de statistique,  Universit\'e de Montr\'eal, \\C.P. 6128 succ. Centre-Ville, Montr\'eal (QC) H3C 3J7, Canada }
\eads{\mailto{post@CRM.UMontreal.CA}, \mailto{wintern@CRM.UMontreal.CA}}
\begin{abstract} In this paper, we derive a nonseparable  quantum superintegrable system in 2D real {E}uclidean space. The Hamiltonian admits no second order integrals of motion but does admit one third and one fourth order integral. We also obtain a classical superintegrable system with the same properties. The quantum system differs from the classical one by corrections proportional to $\hbar^2.$

\end{abstract}

\section{Introduction} 
Let us consider a classical or quantum Hamiltonian given by 
\be \label{H} H=\frac12(p_1^2+p_2^2)+V(x,y)\ee
in two dimensional real Euclidean space, $E_2$. This system is superintegrable if there exist two independent integrals of motion $X$ and $Y$. In classical mechanics $X$ and $Y$ must be well defined functions on phase space and the triplet $( H, X, Y)$ must be functionally independent. The integrals $X$ and $Y$  Poisson commute with $H$ but not with each other. In quantum mechanics $X$ and $Y$ must be well defined (Hermitian) operators in the enveloping algebra of the Heisenberg algebra (or a convergent series in the generators of the Heisenberg algebra) and commute with $H$. They must be algebraically independent within a Jordan algebra generated by $x_j$, $p_j$ and $1.$ Here, $x_1=x,$ $x_2=y$ and the $p_i$'s are their conjugate momenta in the classical system and $p_j=-i\hbar \partial_{x_j}$ in the quantum system.

If $X$ and $Y$ are polynomials of order $n$ and $m$ respectively, (with $n\ge m,$) in the components $p_1,\ p_2 $ of the momenta we shall call the system (\ref{H}) $n$-th order superintegrable. 

All second order superintegrable systems in $E_2$ have been known for some time \cite{FMSUW} as have those in $E_3$ \cite{MSVW1967, Evans1990}. Quite generally quadratic superintegrability is well understood in Euclidean spaces \cite{KMP2}, in spaces of constant curvature \cite{Eis}, and in more general spaces \cite{KKMW2003}. In particular, quadratic superintegrable systems are basically the same in classical and quantum mechanics. The only difference is that in the quantum case operator products have to be symmetrized. Moreover, quadratic integrability in system \eref{H} is related to the separation of variables in the Hamilton-Jacobi and Schr\"odinger  equations, respectively. 
Quadratic superintegrability is hence related to multiseparability. We mention that the best known and most important superintegrable
systems, namely the Kepler-Coulomb system \cite{Fock} and the harmonic oscillator \cite{JH1940} are quadratically superintgrable on $E_n$ for $n\ge 2.$ 

More recently, interest has shifted to higher order superintegrability. Families of superintegrable systems with integrals of motion of arbitrary order have been discovered both in classical and quantum mechanics \cite{TTW2009, TTW2010, PW20101, KMPTTWClass, KKM2010quant, QuesneTTWodd, MPY2010}. A systematic search for superintegrable systems with one second order and one third order integral of motion has be initiated \cite{GW, Gravel, TW20101, MW2008, marquette2009painleve}. The quantum case turns out to be much richer than the classical one. Quantum integrable \cite{Hiet1984, HG1989} and superintegrable \cite{GW, Gravel,TW20101} potentials exist that vanish in the classical limit. In other cases, the classical limit $(\hbar\rightarrow 0)$ is singular and the quantum and classical cases are completely different. 

The purpose of this article is to derive a superintegrable system in $E_2$ that does not allow separation of variables neither in the quantum case not in its classical limit. The integrals of the motion in this case are of order 3 and 4. Previously known superintegrable but nonseparable systems \cite{ADS2006, MPT2010} are purely classical, moreover one of them is complex \cite{MPT2010}

\section{3rd-order integrals}
\subsection{Classical and Quantum Determining Equations}
The determining equations for third order classical and quantum integrals of motion were derived earlier \cite{GW}. They can be presented in both cases in a unified manner. In the quantum case, the integral will have the form 
\be X=\sum_{j+k+l=3}\frac12 A_{jkl}\lbrace L_3^j,p_1^kp_2^l\rbrace +\frac 12\{g_1(x,y), p_1\}+\frac12 \{g_2(x,y),p_2\} \ee
\be L_3=xp_2-yp_1, \qquad p_k=-i\hbar \partial_{x_k}, \quad k=1,2\ee 
where $A_{ijk}$ are real constants and the brackets $\{ \ ,\ \}$ denote anti-commuators. 
We define the polynomials \numparts
\bea   f_1\equiv -A_{300}y^3+A_{210}y^2-A_{102}y^2+A_{030}\\
 f_2\equiv 3A_{300}xy^2-2A_{210}xy+A_{201}y^2+A_{120}x-A_{111}y+A_{021}\\
 f_3\equiv -3A_{300}x^2y-2A_{201}xy+A_{210}x^2+A_{111}x-A_{102}y+A_{012}\\
 f_4\equiv A_{300}x^3+A_{201}x^2+A_{102}x+A_{003}\eea 
\endnumparts 
The commutativity condition $[H, X]=0$ implies 4 determining equations for the three unknown functions $V(x,y), \ g_1(x,y)$ and $ g_2(x,y)$ namely 
 
\bea  \label{g1x}(g_1)_x=3 f_1V_x+ f_2V_y\\
 \label{g2y}(g_2)_y= f_2V_x+3f_4 V_y\\
 \label{g1g2} (g_1)_y+(g_2)_x=2(f_2V_x+f_3V_y)\eea
\bea \label{nonlinear} \fl   g_1V_x+g_2V_y=\\
\fl \quad- \frac{\hbar^2}{4}\left(f_1V_{xxx}+f_2V_{xxy}+f_3V_{xyy}+f_4V_{yyy}+4A_{300}(xV_y-yV_x)+2A_{201}V_x+2A_{201}V_y\right)\nonumber\eea
Equations \eref{g1x}, \eref{g2y}, and \eref{g1g2} are the same in the classical and quantum cases whereas \eref{nonlinear} greatly simplifies for $\hbar \rightarrow 0.$ In both cases  \eref{g1x}, \eref{g2y}, and \eref{g1g2} are linear whereas \eref{nonlinear} is nonlinear. 

The compatibility conditions for the first three determining equations provides a linear partial differential equation (PDE) for the potential 
\bea\fl  0&=&-f_3V_x+(2f_2-3f_4)V_{xxy}+(-3f_1+2f_3)V_{xyy}-f_2V_{yyy}\nn
\fl&&+2(f_{2y}-f_{3y})V_{xx}+2(-3f_{1y}+f_{2x}+f_{3y}-3f_{4x})V_{xy}+2(-f_{2y}+f_{3x})V_{yy}\nn
\label{cc}\fl&&+(-3f_{1yy}+2f_{2xy}-f_{3xx})V_x+(-f_{2yy}+2f_{3xy}-3f_{4xx})V_y.\eea
Further compatibility conditions involving \eref{nonlinear} exist but they are nonlinear and we shall not use them here. 
\subsection{Potentials linear in $y$ that admit a 3rd order integral}
In general the determining equations (\ref{g1x}-\ref{cc})  are difficult to solve. So far, they have been completely solved for potentials $V(x,y)$ that allow separation of variables in Cartesian \cite{Gravel} or polar \cite{TW20101} coordinates. Here, on the contrary, we are looking for potentials that do not allow separation. To do this, we make the Ansatz, 
\be  \label{ansatz}V(x,y) =w_1(x)y+w_0(x), \qquad w_1(x)\ne 0, \qquad w_i:\mathbb{R}\rightarrow \mathbb{R}, \quad i=0,1.\ee
Up to translation in $x$ and $y$ and an irrelevant constant, the only solution of the form \eref{ansatz} of the determining equations in the quantum case is 
\be \label{Vnew} V(x,y)=\frac{\alpha y}{x^{\frac23}}-\frac{5\hbar^2}{72 x^2}, \ee
with $\alpha \in \mathbb{R}.$ 
The quantum integral of motion is 
\be X=3p_1^2p_2+2p_2^3 +\lbrace\frac{9\alpha}{2}x^{\frac13}, p_1\rbrace +\lbrace \frac{3\alpha y}{x^{\frac23}}-\frac{5\hbar^2}{24x^2}, p_2\rbrace\ee  
and the classical potential and integral are obtained by putting $\hbar^2 \rightarrow 0. $ They can of course also be obtained directly by requiring that the Poisson commutator $\lbrace H, X \rbrace_{P}$ vanish. 
\section {Fourth order integrals}
\subsection{Determining equations}
Two basic results on n-th order integrals in $E_2$, valid in both classical and quantum mechanics, are: 
\begin{itemize} 
\item The leading (highest-order in the momenta) term lies in the enveloping algebra of the Euclidean Lie algebra $e(2).$ 
\item All lower order terms in the integral have the same parity as the leading term.\end{itemize} 
It follows that a 4th order integral can be written as 
\be\label{y}\fl \displaystyle Y=\!\!\!\!\!\!\sum_{j+k+l=4} \!\!\!\frac{A_{jkl}}2 \lbrace L_3^j, p_1^kp_2^l\rbrace +\frac12 \left(\lbrace g_1(x,y), p_1^2\rbrace + \lbrace g_2(x,y), p_1p_2\rbrace+ \lbrace g_3(x,y), p_2^2\rbrace\right) +\ell(x,y),\ee
where the $A_{ijk}$ are real constants and $g_i, \ell$ are real functions of $(x,y).$  Since we are not aware of a proof of this result in the literature we provide one for $n=4$ in the Appendix.  The commutation relations $[H,Y]=0$ provides 6 determining equations for the 5 functions $V, g_1 g_2, g_3$ and $\ell$ namely
\bea\label{cq41} g_{1,x}=4f_1V_x+f_2V_y\\
\label{cq42}g_{2,x}+g_{1,y}=3f_2V_x+2f_3V_y\\
\label{cq43}g_{3,x}+g_{2,y}=2f_3V_x+3f_4V_y\\
\label{cq44}g_{2,y}=f_4V_x+4f_5V_y\eea
and 
\bea \label{cq45}\fl\ell_{x}=&2g_1V_x+& g_{2}V_y\\
\fl&+\frac{\hbar^4}{4}\bigg(&(f_2+f_4)V_{xxy}-4(f_1-f_5)V_{xyy}-(f_2-f_6)V_{yyy}\nn
\fl&&+(f_{2,y}-f_{5,x})V_{xx}-(13f_{1,y}+f_{4,x})V_{xy}-4(f_{2,y}-f_{5,x})V_{yy}\nn
\fl && +2(6A_{400}x^2+62A_{400}y^2+3A_{301}x-29A_{310}y+9A_{220}+3A_{202})V_x\nn
\fl && +2(56A_{400}xy+13A_{310}x-13A_{301}y+3A_{211})V_y\bigg)\nn
\label{cq46}\fl \ell_{y}=&g_{2}V_x+&2g_{3}V_y\\
\fl&+\frac{\hbar^2}{4}\bigg(&-(f_2+f_4)V_{xxx}+4(f_1-f_5)V_{xxy}+(f_2-f_6)V_{xyy}\nn
\fl &&+4(f_{1,y}-f_{4,x})V_{xx}-(f_{2,y}+13f_{5,x})V_{xy}-(f_{1,y}-3f_{4,x})V_{yy}\nn
\fl &&+2(56A_{400}xy-13A_{310}x+13A_{301}y+3A_{211})V_x\nn
\fl &&+2(62A_{400}x^2+6A_{400}y^2+29A_{A301}x-3A_{310}y+9A_{202}+3A_{220})V_y\bigg).\nonumber\eea

The polynomials $f_i$ are defined as
\numparts
\bea
\fl f_1&=A_{400}y^4-A_{310}y^3+A_{220}y^2-A_{130}y+A_{040}\\
\fl f_2&=-4A_{400}xy^3-A_{301}y^3+3A_{310}xy^2+A_{211}y^2-2A_{220}xy-A_{121}y+A_{130}x+A_{031}\\
\fl f_3&=6A_{400}x^2y^2+3A_{301}xy^2-3A_{310}x^2y+A_{202}y^2+A_{220}x^2-A_{112}y-A_{121}x+A_{022}\\
\fl f_4&=-4A_{400}yx^3+A_{310}x^3-3A_{301}x^2y+A_{211}x^2-2A_{202}xy+A_{112}x-A_{103}y+A_{013}\\
\fl f_5&=A_{400}x^4+A_{301}x^3+A_{202}x^2+A_{103}x+A_{004}
\eea
\endnumparts
The compatibility condition for \eref{cq41}, \eref{cq42}, \eref{cq43}, and \eref{cq44}  is a fourth order linear PDE for $V$ given by 
\[0= \partial_{yyy}\left(4f_1V_x+f_2V_y \right)-\partial_{xyy}\left(3f_2V_x+2f_3V_y \right) \]
\be \label{cc4} \quad +\partial_{xxy}\left(2f_3V_x+3f_4V_y \right)-\partial_{xxx}\left(f_4V_x+4f_5V_y \right).\ee

Equations \eref{cq45} and \eref{cq46} have a compatibilty condition which is a non-linear PDE for $V, g_1, g_2, g_3.$ 
Again, the classical determining equations are obtained in the limit $\hbar\rightarrow 0.$ Thus,  \eref{cq41}, \eref{cq42}, \eref{cq43},\eref{cq44} and hence \eref{cc4} are the same in classical mechanics but  \eref{cq45} and \eref{cq46} have a quantum correction of order $\hbar^2.$ 

\subsection{A superintegrable nonseparable system}
It is now quite easy to check that the potential \eref{Vnew} satisfies the quantum and classical determining equations for the existence of a fourth order integral. Finally, we have obtained the main results of this article. 

\begin{theorem} The operator triplet $( H, X, Y)$ with 
\be \label{hq} H=\frac12(p_1^2+p_2^2)+\frac{\alpha y}{x^{\frac23}}-\frac{5\hbar^2}{72x^2}\ee
 \be\label{xq} X=3p_1^2p_2+2p_2^3 +\lbrace\frac{9\alpha}{2}x^{\frac13}, p_1\rbrace +\lbrace \frac{3\alpha y}{x^{\frac23}}-\frac{5\hbar^2}{24x^2}, p_2\rbrace\ee
 \be \label{yq} \fl Y=p_1^4 +\left\{\frac{2\alpha y}{x^{\frac23}}-\frac{5\hbar^2}{36 x^2},p_1^2\right\}-\left\{6x^{\frac13}\alpha, p_1p_2\right\}-\frac{2\alpha^2(9x^2-2y^2)}{x^{\frac43}}  -\frac{5\alpha\hbar^2 y}{9x^{\frac83}}+\frac{25\hbar^4}{1296x^4}\ee
 constitutes a quantum superintegrable system that does not allow multiplicative separation of variables in the Schr\"odinger equation in any system of coordinates. 
\end{theorem}

\begin{theorem} The triplet of well defined functions on phase space with 

 \be  \label{hc} H=\frac12(p_1^2+p_2^2)+\frac{\alpha y}{x^{\frac23}}\ee
 \be\label{xc} X=3p_1^2p_2+2p_2^3 +9\alpha x^{\frac13} p_1 +\frac{6\alpha y}{x^{\frac23}} p_2\ee
 \be \label{yc} Y=p_1^4 +\frac{4\alpha y}{x^{\frac23}}p_1^2-12x^{\frac13}\alpha p_1p_2-\frac{2\alpha^2(9x^2-2y^2)}{x^{\frac43}} \ee
 constitutes a classical superintegrable system that does not allow additive separation of variables in the Hamilton-Jacobi equation in any system of coordinates. 
 \end{theorem}
 Both of these theorems are proved by directly verifying that they satisfy the determining equations. The nonseparability result follows from the fact that the Hamiltonian does not allow any second order integrals of motion. 
 
 In the classical case it is easy to verify that $( H, X, Y )$ of (\ref{hc}), \eref{xc} and \eref{yc} are functionally independent. Indeed, the Jacobian matrix $J$ satisfies
 \be J=\frac{\partial (H,X,Y)}{\partial x, y, p_1, p_2}, \qquad \mbox{ rank } J=3. \ee
 Hence, no nontrivial relations of the type $0=F(X, Y, H)$ exist (in particular there is no syzygy). In the quantum case, if a Jordan polynomial relation between the three integrals \eref{xq}, \eref{yq} and \eref{hq} did exist, it would imply the existence of a syzygy in the classical limit $\hbar \rightarrow 0.$ 
 
The integrals of motion in both cases generate a finite dimensional decomposable Lie algebra 
\be \label{algebra} \lbrace X, Y, \mathbb{I} \rbrace \oplus H\ee
with 
\be [X,Y]=-108\alpha^3\mathbb{I}.\ee
The algebra \eref{algebra} is thus a direct sum of a Heisenberg algebra with $H$ as an additional central element.

 \section{Conclusion}
 To our knowledge, \eref{Vnew} is the first quantum nonseparable superintegrable system in the literature. Classical ones, on the other hand, are already known. To our knowledge, the first known one is the classical nonperiodic N particle Toda lattice. M Agrotis et al \cite{ADS2006} have shown that the $N$ particle Toda system allows $2N-1$ integrals. A different classical nonseparable superintegrable system in complex two dimensional Euclidean space was recently presented by Maciejewski et al \cite{MPT2010}. 
 
 The first systematic search for integrable systems with a third order integral in two dimensions was published by Drach in 1935 \cite{Drach} and was conducted in a two dimensional complex space $E_2(\mathbb{C}).$ He found 10 families of potentials. One of them can be rewritten in real Euclidean space $E_2$ as 
 \be \label{VH} V=\frac1{x^{\frac23}}\left(a+by+c(4x^2+3y^2)\right).\ee
 We see that the superintegrable potential \eref{hc} is a special case of \eref{VH} (the constant $a$ can be translated away for $b\ne 0$). For $c\ne 0$ \eref{VH} does not allow a fourth order integral, though it still might be superintegrable.

 \ack{The research of P. W. is partially supported by a research grant from NSERC of Canada. S. P. acknowledges a postdoctoral fellowship awarded by the Laboratory of Mathematical Physics of the Centre de Recherches Math\'ematiques, Universit\'e de Montr\'eal. }
 \appendix
 \setcounter{section}{1}
 \section*{Appendix: Proof of equation \eref{y}}
\begin{theorem} Given a self-adjoint Hamiltonian of the form 
\be H=\frac12(p_1^2+p_2^2)+V(x,y), \qquad V:\mathbb{R}^2\rightarrow \mathbb{R}.\ee
Any fourth-order integral of motion, can be written in the form \eref{y} modulo an all over complex constant and other integrals of motion.   
\end{theorem}
{\bf Proof:} First, since $H$ is a real differential operator, the real and imaginary parts of $Y$ must commute with $H$ separately and so $Y$ can be assumed real and can be written as
\be Y=\sum_{k=0}^{4}\sum_{j=0}^{j} f_{j k} p_1^{j}p_2^{k-j}, \quad p_j=-i\hbar \partial_{x^j} \ee
with $f_{j 2\ell}:\mathbb{R}^2\rightarrow \mathbb{R}$ and $f_{j 2\ell+1}:\mathbb{R}^2\rightarrow i \mathbb{R}.$

Next,  note that $H$ is self-adjoint and so any integral which commutes with it can be broken up into self-adjoint and skew-adjoint parts which will simultaneously commute with $H$. Further,  at least one of the parts will remain fourth order and so, up to multiplication by a complex constant and modulo lower order terms, the constant of the motion can be assumed to be self-adjoint and can be written as 
\be Y=\frac12(Y^\dagger +Y)\ee
which implies
\be \fl Y=\frac 12 \left( \lbrace f_{j4}, p_1^jp_2^{4-j}\rbrace + \lbrace f_{j2}, p_1^jp_2^{2-j}\rbrace+f_{0,0} +[f_{j3}, p_1^jp_2^{3-j}]+[f_{j1},p_1^jp_2^{1-j}]\right).\ee

However,  the odd terms can be written as symmetrized even terms. Computing directly, 

\be [f_{11}, p_1]=i\hbar f_{11,x}\qquad  [f_{01}, p_2]=i\hbar f_{12,y}\ee
\bea [f_{33}, p_1^3]&=i\hbar^3\left( -f_{33, xxx} -3f_{33,xx}\partial_x-3f_{33,x}\partial_x^2\right)\nn
&= i\hbar^3\left(-f_{33,xxx}-\frac{3}2\lbrace f_{33,x}, \partial_x^2\rbrace +\frac{3}2f_{33,xxx}\right)\nn
&=\frac{i\hbar^3}{2}f_{33,xxx}+\frac{3i\hbar}{2}\lbrace f_{33,x},p_1^2\rbrace, \eea
\bea  [f_{23}, p_1^2p_2]&=i\hbar^3\left(-f_{23,xxy}-2f_{23,xy}\partial_x-f_{23,xx}\partial_y-2f_{23,x}\partial_x\partial_y-f_{23,y}\partial^2_x\right)\nn
&=i\hbar^3\left(-\frac12 \lbrace f_{23,y},\partial_x^2\rbrace -\lbrace f_{23,x}, \partial_x\partial_y\rbrace +\frac12 f_{23,xxy}\right)\nn
&=\frac{i\hbar} 2 \lbrace f_{23,y},p_1^2\rbrace +i\hbar \lbrace f_{23,x},p_1p_2\rbrace +\frac{i\hbar^3}2 f_{23,xxy}\eea
and similarly 
\bea  [f_{13}, p_1p_2^2]&=\frac{i\hbar} 2 \lbrace f_{13,x},p_2^2\rbrace +i\hbar \lbrace f_{13,y},p_1p_2\rbrace +\frac{i\hbar^3}2 f_{13,xyy}\\
 \left[ f_{03}, p_2^3\right] &=\frac{i\hbar^3}{2} f_{03,yyy}+\frac{3 i\hbar }{2}\lbrace f_{03,y},p_2^2 \rbrace. \eea
Here we recall that the the odd order functions are purely imaginary and so we can define new, real functions 
\bea a_{02}&=f_{22}+\frac{3i\hbar}{2}f_{33,x}  +\frac{i\hbar} 2f_{23,y} \\
        a_{12}&=f_{12}+i\hbar f_{23,x}+i\hbar f_{13,y}\\
        a_{22}&=f_{02}+\frac{i\hbar} 2 f_{13,x}+\frac{3 i\hbar }{2} f_{03,y}\eea
        
      \be \fl  b=f_{0,0}+ \frac{i\hbar^3}{2}f_{33,xxx}+ \frac{i\hbar^3}2 f_{23,xxy}+\frac{i\hbar^3}2 f_{13,xyy}+\frac{i\hbar^3}{2} f_{03,yyy} +i\hbar f_{11,x}+i\hbar f_{12,y}       \ee

To finish the theorem, we need only show that the highest-order terms lies in the enveloping algebra of the Euclidean Lie algebra $e(2).$ If we compute the determining equations that result from the requirement that $[Y,H]=0,$ the differential equations that $f_{j4}$ must solve are identical to those determined by the requirement that 
\[ \left[\sum_{j=0}^4\frac12 \lbrace f_{j4}, p_1^jp_2^4\rbrace,\partial_x^2+\partial_y^2\right] =0. \] Thus,  there exist constants $A_{ijk}$ such that the two operators  $\frac12 \lbrace f_{j4}, p_1^jp_2^4\rbrace$ and $A_{jk\ell} \lbrace L_3^j,p_1^kp_2^\ell\rbrace$ differ by at most lower order terms. However, since both of these operators are real and self-adjoint their difference must also be. Thus, the difference between the two operators is at most an order two differential operator which has only even terms in symmetrized form, as proven above. That is there exist real functions $c_{j2}$, $d$ such that 
\be \sum_{j=0}^4\frac12 \lbrace f_{j4}, p_1^jp_2^4\rbrace-\sum_{j+k+l=4} A_{jk\ell} \lbrace L_3^j,p_1^kp_2^\ell\rbrace =\sum_{j=0}^2 \frac12 \lbrace c_{j2}, p_1^j p_2^{2-j}\rbrace +d.\ee
Finally, if we define 
\be g_1=a_{02}+c_{02}, \quad g_2=a_{12}+c_{12}, \quad g_3=a_{22}+c_{22}\quad \ell=b+d\ee
then, $Y$ is of the form \eref{y}. \begin{flushright} $\square$\end{flushright}

 \section*{References}
 \bibliography{all}
 \bibliographystyle{jphysa}
  \end{document}